\documentclass[
    ,final
  ]
  {aipproc}

\layoutstyle{6x9}
\begin{document}

\title{The Explicit Procedures for Reconstruction of Full Set of Helicity
Amplitudes in Elastic Proton-Proton and Proton-Antiproton
Collisions}


\author{V.A. Okorokov}{
  address={Moscow Engineering Physics Institute (State University), 115409,
Moscow,
  Russia} }

\author{S.B. Nurushev}{
  address={Institute for High Energy Physics, 142281
Protvino, Moscow Region, Russia} }

\begin{abstract}
The explicit procedures are described for reconstruction of the
full set of helicity amplitudes in proton-proton and
proton-antiproton elastic scattering. The procedures are based on
the derivative relations for the helicity amplitudes in
$s$-channel, on the explicit parametrization of the leading spin
non flip amplitudes and crossing - symmetry relations. Asymptotic
theorems are used for definition of free parameters in derivative
relations. We also study the Odderon influence on the helicity
amplitude reconstruction. Reconstruction procedures are valid at
extremely wide energy domain and broad range of momentum transfer.
These procedures might be useful in studying the spin phenomena in
proton-proton and proton-antiproton elastic scattering.
 \end{abstract}

\maketitle


\section{Introduction}

    Elastic scattering of hadrons has always been a crucial tool for
study the dynamics of strong interaction. In the absence of the
strong interaction  theory the predictions or interpretation of
the experimental observables are furnished by the phenomenological
approaches. One of the reliable approaches is the asymptotical
model. Other models (like Regge model) suffer from necessity to
introduce many free parameters which should be defined by fitting
to the experimental data. In such a situation the direct
reconstruction of the scattering matrix from the complete set of
the experimental data will be the appropriate method.
Unfortunately such set of experiments were never been fulfilled in
the high energy region. One may hope that such program will be
realized at RHIC, FAIR (PAX project) and other facilities in near
future. In order to make the predictions for the measurable
observables in those facilities one needs to have a method which
should be well justified, contains a small number of free
parameters and applicable at wide range of the kinematics
variables. Below we propose such a technique. Its applicability
may be tested by the joint consideration of proton-proton and
proton-antiproton elastic scattering data.

\section{Reconstruction procedure}

Two methods have been suggested for building  the full set of
helicity amplitudes for elastic $p\bar{p}$-collisions in
\cite{Okorokov-2001, Okorokov-2005}. We describe in details the
method for deducing the  helicity amplitudes based on
crossing-symmetry relation and the derivative relations here.The
amplitudes for the binary reaction $A+B \to C+D$ in
$s-,~t-,~\mbox{and}~u-$ channels all depend upon the Mandelstam
variables and are described by just one set of analytic functions
evaluated in different regions of variables $s,t,u$. Thus the
following preliminary expression for full set of helicity
amplitudes of $p\bar{p}$ elastic scattering via set of helicity
amplitudes for $pp$ elastic reaction have been derived in
\cite{Okorokov-2005}:
\begin{equation}
\left. \begin{array}{l} \vspace{0.2cm}
 \Phi _1^{p\bar p}  = {1 \mathord{\left/
 {\vphantom {1 2}} \right.
 \kern-\nulldelimiterspace} 2}\,\left[ {\sin ^2 \psi \left( {\Phi _1^{pp}
+ \Phi _2^{pp}  + \Phi _3^{pp} } \right)\, + \left( {1 + \cos ^2
\psi } \right)\Phi _4^{pp} } \right]
 \\
\vspace{0.2cm}
 \Phi _2^{p\bar p}  = {1 \mathord{\left/
 {\vphantom {1 2}} \right.
 \kern-\nulldelimiterspace} 2}\,\left[ {\sin ^2 \psi \left( {\Phi _1^{pp}
+ \Phi _3^{pp}  - \Phi _4^{pp} } \right) - \left( {1 + \cos ^2
\psi }
\right)\Phi _2^{pp} } \right]\, \\
\vspace{0.2cm}
 \Phi _3^{p\bar p}  = {1 \mathord{\left/
 {\vphantom {1 2}} \right.
 \kern-\nulldelimiterspace} 2}\,\left[ {\sin ^2 \psi \left( {\Phi _1^{pp}
+ \Phi _2^{pp}  - \Phi _4^{pp} } \right) - \left( {1 + \cos ^2
\psi }
\right)\Phi _3^{pp} } \right] \\
\vspace{0.2cm}
 \Phi _4^{p\bar p}  = {1 \mathord{\left/
 {\vphantom {1 2}} \right.
 \kern-\nulldelimiterspace} 2}\,\left[ {\left( {1 + \cos ^2 \psi }
\right)\Phi _1^{pp}  - \sin ^2 \psi \left( {\Phi _3^{pp}  + \Phi
_2^{pp}  -
\Phi _4^{pp} } \right)} \right] + 2\Phi _5^{pp} \sin \psi  \\
\vspace{0.2cm}
 \Phi _5^{p\bar p}  = {1 \mathord{\left/
 {\vphantom {1 2}} \right.
 \kern-\nulldelimiterspace} 2}\cos \psi \left[ {\sin \psi \left( {\Phi
_1^{pp}  + \Phi _2^{pp}  + \Phi _3^{pp}  - \Phi _4^{pp} } \right)
+ 2\Phi
_5^{pp} } \right] \\
 \end{array} \right\}\label{helamp-pantip-sys}
\end{equation}
where
$$\cos \psi  =  \sqrt {\frac{\textstyle st}{\textstyle \left(s -
4m_{p}^{2} \right)\left(t - 4m_{p}^{2}\right)}};~\sin \psi  =
\frac{\textstyle m_{p} \sqrt {s - 4m_{p}^{2}}}{\textstyle \sqrt
{t\left(t - 4m_{p}^{2}\right)}}\sin \theta ;~\cos \theta  = 1 +
\frac{\textstyle 2t}{\textstyle s - 4m_{p}^{2}};$$ $\theta$ - CM
scattering angle in the $s$-channel, $m_{p}$ - proton mass. The
system \eqref{helamp-pantip-sys} shows apparent analytical forms
for full set of amplitudes of elastic $p\bar{p}$ scattering
$\left\{ {\Phi _{i}^{p\bar p} } \right\}_{i = 1-6}$ via helicity
amplitudes $\left\{ {\Phi _{j}^{pp} } \right\}_{j = 1-5}$ for
crossing-symmetrical $pp$ channel. Lets to stress that the
relation $\Phi _{5}^{pp}=-\Phi _{6}^{pp}$ has been taken into
account in the \eqref{helamp-pantip-sys} already. The relation
$\left|\Phi _{5}^{p\bar p}\right|=\left|\Phi _{6}^{p\bar
p}\right|$ corrects according to $G$-parity conservation in
elastic $p\bar{p}$ collisions \cite{Goldberger-1960}. But on the
other hand the exact correlation between $\left|\Phi _{5}^{p\bar
p}\right|$ and $\left|\Phi _{6}^{p\bar p}\right|$ is open question
in general case because of Odderon pole contribution is still a
contentious topic.

\subsection{Derivative relations}

The $pp$ elastic scattering under study in order to obtain some
additional correlations for set of helicity amplitudes $\left\{
{\Phi _{j}^{pp} } \right\}_{j = 1-5}$. Usually the following
formula is suggested for spin non-flip amplitudes
$\Phi_{1}^{pp}=\Phi_{3}^{pp}$. The derivative relations allow to
express spin-flip amplitude $\Phi_{5}^{pp}$ and spin double-flip
amplitude $\Phi_{4}^{pp}$ via $\Phi_{1}^{pp}$
\cite{Schrempp-1975}:
\begin{equation}
\Phi_{5}^{pp}\left(s,t\right)=C_{1}^{pp}(s)\frac{\textstyle
\partial}{\textstyle \partial(\sqrt{-t})}\Phi_{1}^{pp}\left(s,t\right);~~
\Phi_{4}^{pp}\left(s,t\right)=C_{2}^{pp}(s)\frac{\textstyle
\partial^{2}}{\textstyle
\partial(\sqrt{-t})^{2}}\Phi_{1}^{pp}\left(s,t\right),
\label{derriv-rel-my}
\end{equation}
where $C_{k}^{pp}(s)=C_{k1}^{pp}(s)+iC_{k2}^{pp}(s),~k=1,2$ -
complex parameters in general. The some versions of additional
correlation for spin double-flip helicity amplitude
$\Phi_{2}^{pp}$ were discussed in \cite{Okorokov-2001,
Okorokov-2005, Leader-2001}. We would like to emphasize that
complex parameters $C_{k}(s) (k=1,2)$ must be defined for exact
knowledge of spin-flip and double-spin amplitudes for $pp$ elastic
scattering.

\subsection{Determination of the free parameters}

It seems that the combination of sets of helicity amplitudes for
$pp$ and $p\bar{p}$ elastic reactions namely in the framework of
the method described above and some additional equations result in
several ways for analytic determination of parameters $C_{k}^{pp}
(k=1,2)$. The Odderon hypothesis is crucial important for
definition of unknown parameters $C_{k}^{pp} (k=1,2)$ in the
derivative relations \eqref{derriv-rel-my}. We suggest to use the
asymptotic behavior of total cross section, differential cross
section, $\rho$ and $B$ parameters in order to obtain the complex
unknown parameters in high-energy domain.

The most general case corresponds to possibility for Odderon
exchange as well as for Pomeron one. The general Pomeranchuk
theorem is satisfied in framework Odderon hypothesis, but the
original Pomeranchuk theorem is violated. Thus one can get the
following equation system for definition of $C_{k}^{pp} (k=1,2)$
parameters:
\begin{equation}
\left. \begin{array}{l}
 \Delta\sigma_{\mbox{tot}} \equiv
\sigma_{\mbox{tot}}^{p\bar{p}}-\sigma_{\mbox{tot}}^{pp}
 \propto \left[\mbox{Im}\Phi_{1}^{p\bar{p}}\left(s,t=0\right)-
 \mbox{Im}\Phi_{1}^{pp}\left(s,t=0\right)\right],
 \\
\Delta \rho =\rho^{p\bar{p}}-\rho^{pp}, \\
\Delta  \left(d\sigma_{\mbox{el}}/dt\right)=
\left(d\sigma_{\mbox{el}}/dt\right)^{p\bar{p}}-
\left(d\sigma_{\mbox{el}}/dt\right)^{pp}, \\
 \Delta B =B^{p\bar{p}}-B^{pp}.
 \end{array} \right\}\label{eq-sys-1-odderon}
\end{equation}
The definition of the parameters $C_{k}^{pp} (k=1,2)$ becomes
model dependent and non trivial task because of model dependent
values are on the left parts of equation in the system
\eqref{eq-sys-1-odderon}. According to accelerator and cosmic ray
experimental data \cite{Avila-2003} and phenomenological estimates
\cite{Block-2006} also the values of
$\sigma_{\mbox{tot}}^{p\bar{p}}$ are (very) close to corresponding
values of $\sigma_{\mbox{tot}}^{pp}$ up to energy $\sqrt{s}=100$
TeV. The usual interpretation of the experimental data for
$\rho$-parameter is that $\Delta \rho$ is very small. It should be
emphasized that even $\Delta \rho=0$ would not exclude an Odderon,
but would only rule out specific models for the soft Odderon. Thus
one can suggests $\Delta \sigma_{\mbox{tot}}=0$ and $\Delta
\rho=0$ without full excluding of Odderon pole contribution and
without significant violation of general description consequently.
One can use only two first equation in the system
\eqref{eq-sys-1-odderon} in the framework of simple suggestions
that there is one complex parameter or two clear real / imagine
parameters. It should be emphasized that the unknown parameters
would be define model independently in these specific cases even
taking into account possible presence of the (soft) Odderon
contribution.

As shown above the new experimental date and phenomenological
investigations will be very important at ultra-high energy
$\sqrt{s} \sim 100$ TeV in particulary for decision some
fundamental problems for hadron interactions and distinguishing
different theoretical models.

The original Pomerahcuk theorems, namely, for total cross section
and for differential cross section in binary reaction, and
Cornille-Martin theorem for the forward slope parameter $B$
\cite{Barone-2003} can be used in the framework of hypothesis for
presence only Pomeron exchange. The system for definition of free
parameters $C_{k}^{pp} (k=1,2)$ is given by
\begin{equation}
\left. \begin{array}{l}
 \mbox{Im}\Phi_{1}^{p\bar{p}}\left(s,t=0\right) =
 \mbox{Im}\Phi_{1}^{pp}\left(s,t=0\right),
 \\
\mbox{Re}\Phi_{1}^{p\bar{p}}\left(s,t=0\right) =
 \mbox{Re}\Phi_{1}^{pp}\left(s,t=0\right),
 \\
\left(d\sigma_{\mbox{el}}/dt\right)^{p\bar{p}}=
\left(d\sigma_{\mbox{el}}/dt\right)^{pp}, \\
 B^{p\bar{p}}=B^{pp}.
 \end{array} \right\}\label{eq-sys-1-pomeron}
\end{equation}

This system allows to determine all components of complex free
parameters by model independent way. Moreover for Pomeron exchange
only there is additional correlation between helicity amplitudes
for $p\bar{p}$ elastic reaction, namely, $\left|\Phi _{5}^{p\bar
p}\right|=\left|\Phi _{6}^{p\bar p}\right|$ just as well as for
$pp$ scattering. The asymptotic relations and fit of experimental
dependences at low and intermediate energies should be used for
determination of free parameters for the reconstruction method
understudy.

\section{Summary}

The main results of this paper are following. We suggest the
explicit procedure for the reconstruction of the the full set of
the helicity amplitudes for the elastic $p\bar{p}$ scattering.
This method is based on fundamental crossing-symmetry property,
derivative relations for helicity amplitudes and selection the
anatlytical expression for the spin non-flip amplitude describing
well the $pp$ experimental data. We apply to this spin non-flip
amplitude the derivative procedure for finding all $pp$ helicity
amplitudes. After proving that we get the good description of
$pp$-data we apply the crossing relations in order to find the
helicity amplitudes for the $p\bar{p}$ elastic scattering. We
introduce the minimum of the free parameters which are fixed
through the asymptotic relations taking into account the Pomeron
and Odderon contributions. The unified analysis of the $pp$ and
$p\bar{p}$ data allows to check in details the proposed method. It
seems this analytical method might be useful for direct
reconstruction of elements of the scattering matrix at extremely
wide initial energy domain and broad range of momentum transfer.

\end{document}